\documentclass[aps,twocolumn,pre,showpacs,10pt]{revtex4-1} 
\usepackage{setspace}
\usepackage{graphicx}
\usepackage{subfigure}
\usepackage{epsfig}
\usepackage{amsmath}
\usepackage{epsfig}
\usepackage{amssymb}

\begin{document}

\title{Quantum quenches in the thermodynamic limit. II. Initial ground gtates}

\author{Marcos Rigol}
\affiliation{Department of Physics, The Pennsylvania State University,
University Park, Pennsylvania 16802, USA}

\begin{abstract}
A numerical linked-cluster algorithm was recently introduced to study quantum quenches 
in the thermodynamic limit starting from thermal initial states 
[M. Rigol, Phys. Rev. Lett. {\bf 112}, 170601 (2014)]. Here, we tailor that algorithm
to quenches starting from ground states. In particular, we study quenches from the 
ground state of the antiferromagnetic Ising model to the $XXZ$ chain. Our results for 
spin correlations are shown to be in excellent agreement with recent analytical 
calculations based on the quench action method. We also show that they are different
from the correlations in thermal equilibrium, which confirms the expectation that 
thermalization does not occur in general in integrable models even if they cannot 
be mapped to noninteracting ones.
\end{abstract}

\pacs{03.75.Kk, 03.75.Hh, 05.30.Jp, 02.30.Ik}

\maketitle

Interest in the far-from-equilibrium dynamics of isolated quantum systems is on the rise 
\cite{rigol_dunjko_08_34,cazalilla_rigol_10_46,dziarmaga_10,polkovnikov_sengupta_review_11}. 
Among the questions that are currently being addressed are 
\cite{rigol_dunjko_08_34,cazalilla_rigol_10_46,dziarmaga_10,polkovnikov_sengupta_review_11}: 
(i) How do observables evolve and equilibrate in isolated systems far from equilibrium? 
(ii) How can one determine expectation values of observables after equilibration (if it occurs)? 
(iii) Do equilibrated values of observables admit a statistical mechanics description?
(iv) Is the relaxation dynamics and description of observables after relaxation different 
in integrable and nonintegrable systems? In this work we address questions
(ii)--(iv) in the context of quantum quenches. 

We start with a system characterized by an initial 
density matrix $\hat{\rho}^I$ (which is stationary under an initial Hamiltonian $\hat{H}_I$) 
and study the result of its time evolution under unitary dynamics dictated by 
$\hat{H}$, $\hat{\rho}(\tau)=\exp[-\imath\hat{H}\tau/\hbar]\hat{\rho}^I\exp[\imath\hat{H}\tau/\hbar]$,
where $\tau$ denotes time. We assume that $\hat{\rho}^I$ is not stationary under $\hat{H}$.
As discussed in numerical \cite{rigol_dunjko_08_34,rigol_09_39,rigol_09_43,gramsch_rigol_12_78,
ziraldo_silva_12,he_santos_13_84,ziraldo_santoro_13,zangara_dente_13,sorg_vidmar_14} and analytical 
\cite{cramer_dawson_08,barthel_schollwock_08,reimann_08,linden_popescu_09,cramer_eisert_10,
gogolin_muller_11,campos_zanardi_13} studies, if an observable $\hat{O}$ 
equilibrates, its expectation value after equilibration can be computed as 
$\hat{O}^\text{DE}=\text{Tr}[\hat{\rho}^{\text{DE}}\hat{O}]$. 
$\hat{\rho}^{\text{DE}}\equiv\text{lim}_{\tau'\rightarrow\infty}1/\tau'\int_0^{\tau'} 
d\tau\,\hat{\rho}(\tau)=\sum_\alpha W_\alpha |\alpha\rangle\langle\alpha|$ is the density 
matrix in the so-called diagonal ensemble (DE) \cite{rigol_dunjko_08_34} and $W_\alpha$ 
are the diagonal matrix elements of $\hat{\rho}^I$ in the basis of the eigenstates 
$|\alpha\rangle$ of $\hat{H}$, which are assumed to be nondegenerate. For initial 
thermal states, $\hat{O}^\text{DE}$ can be computed using numerical linked-cluster expansions 
(NLCEs) as discussed in Ref.~\cite{rigol_14_91}. Here we show how to use NLCEs when the initial 
state is a ground state. 

Linked-cluster expansions \cite{oitmaa_hamer_book_06} allow one to compute expectation values 
of extensive observables (per lattice site, $\mathcal{O}$) in translationally invariant lattice 
systems in the thermodynamic limit. This is done by summing over the contributions from 
all connected clusters $c$ that can be embedded on the lattice
\begin{equation}
\label{eq:LCE1}
\mathcal{O}=\sum_{c}M(c)\times \mathcal{W}_{\mathcal{O}}(c),
\end{equation}
where $M(c)$ is the multiplicity of $c$ (number of ways per site in which $c$ can be embedded 
on the lattice) and $\mathcal{W}_{\mathcal{O}}(c)$ is the weight of a given observable $\hat{O}$ in $c$. 
$\mathcal{W}_{\mathcal{O}}(c)$ is calculated using the inclusion-exclusion principle:
\begin{equation}
\label{eq:LCE2}
 \mathcal{W}_{\mathcal{O}}(c)=\mathcal{O}(c)-\sum_{s \subset c} \mathcal{W}_{\mathcal{O}}(s).
\end{equation}
In Eq.~\eqref{eq:LCE2}, the sum runs over all connected sub-clusters of $c$ and
\begin{equation}
\label{eq:LCE3}
\mathcal{O}(c)={\textrm{Tr} [\hat{\mathcal{O}}\,\hat{\rho}_c]}/
{\textrm{Tr} [\hat{\rho}_c]}
\end{equation}
is the expectation value of $\hat{\mathcal{O}}$ calculated for the finite cluster $c$,
with the many-body density matrix $\hat{\rho}_c$. In thermal equilibrium, linked-cluster 
calculations are usually implemented in the grand-canonical ensemble (GE), so
$\hat{\rho}_c\equiv\hat{\rho}^\text{GE}_c=e^{-(\hat{H}_c-\mu \hat{N}_c)/k_B T}/
\textrm{Tr} [e^{-(\hat{H}_c-\mu \hat{N}_c)/k_B T}]$. $\hat{H}_c$ and $\hat{N}_c$ 
are the Hamiltonian and the total particle number operators in cluster $c$, 
$\mu$ and $T$ are the chemical potential and the temperature, respectively, and 
$k_B$ is the Boltzmann constant ($k_B$ is set to unity in what follows).

Within NLCEs, $\mathcal{O}(c)$ in Eq.~\eqref{eq:LCE3} is calculated using exact 
diagonalization \cite{rigol_bryant_06_25,rigol_bryant_07_30,rigol_bryant_07_31} 
(for a pedagogical introduction to numerical linked-cluster expansions and their 
implementation, see Ref.~\cite{tang_khatami_13}). 
For various lattice models of interest in thermal equilibrium, NLCEs typically converge 
at lower temperatures than high-temperature expansions 
\cite{rigol_bryant_06_25,rigol_bryant_07_30,rigol_bryant_07_31}. 
In order to use NLCEs to make calculations in the DE after a quench 
starting from a thermal state \cite{rigol_14_91}, the system is assumed to be disconnected from 
the bath at the time of the quench, at which, in each cluster $c$,
$\hat{H}^I_c\rightarrow\hat{H}_c$. One can then write the density matrix of
the DE in each cluster as
\begin{equation}\label{eq:rhode}
 \hat{\rho}^\text{DE}_c=\sum_\alpha W^c_\alpha|\alpha_c\rangle \langle\alpha_c|,
\end{equation}
where $W_\alpha^c =(\sum_a e^{-(E^c_a-\mu_I N^c_a)/{T_I}}|\langle\alpha_c|a_c\rangle|^2)/Z^I_c$, 
$|\alpha_c\rangle$ are the eigenstates of $\hat{H}_c$,
$|a_c\rangle$ ($E_a^c$) are the eigenstates (eigenvalues) of $\hat{H}^I_c$, $N_a^c$ is 
the number of particles in $|a_c\rangle$, $\mu_I$, $T_I$, and 
$Z^I_c=\sum_a e^{-(E^c_a-\mu_I N_a^c)/{T_I}}$ are the initial chemical potential, temperature, 
and partition function, respectively. Using $\hat{\rho}^\text{DE}_c$ instead of 
$\hat{\rho}^\text{GE}_c$ in the calculation of $\mathcal{O}(c)$, NLCEs can be used to 
compute observables in the DE after a quench in the thermodynamic limit \cite{rigol_14_91}.

For initial Hamiltonians in which correlations are short ranged at all temperatures, one can, in 
principle, use NLCEs as described to compute $\hat{O}^\text{DE}$ after a quench 
starting from the ground state. The idea would be to take $T_I$ to be low enough so 
that the initial state is essentially the ground state of the system. For equilibrium
properties, this was shown to work for two-dimensional lattice systems in 
Refs.~\cite{rigol_bryant_06_25,rigol_bryant_07_30}. However, it is much more efficient
to implement a NLCE only considering the ground state. The latter can be calculated,
e.g., using the Lanczos algorithm \cite{khatami_singh_11_66}, without the need of fully 
diagonalizing the Hamiltonian. Furthermore, if one is interested in quenches from known 
initial states, then there is no need to perform any diagonalization at all.

In order to discuss how NLCEs can be implemented for initial ground states, or potentially 
any pure state, we focus on the ground state of the antiferromagnetic (AF) 
Ising chain as the initial state, and consider quenches to the (integrable) $XXZ$ chain
\cite{cazalilla_citro_review_11_64} with
\begin{equation}\label{eq:XXZ}
 \hat{H}=J\left(\sum_i \sigma_i^x \sigma_{i+1}^x + \sigma_i^y \sigma_{i+1}^y + 
         \Delta  \sigma_i^z \sigma_{i+1}^z\right),
\end{equation}
where $\sigma^x$, $\sigma^y$, and $\sigma^z$ are the Pauli matrices, we set $J$ (and $\hbar$) 
to unity, and $\Delta$($\geq1$) is the anisotropy parameter.
The ground state of the AF Ising chain is degenerate, 
$|\uparrow\downarrow\uparrow\downarrow\ldots \rangle$ and
$|\downarrow\uparrow\downarrow\uparrow\ldots \rangle$. Their even and odd superposition, 
which preserve translational invariance in the thermodynamic limit, are the ones that enter 
in the NLCE. This follows from the fact that, to diagonalize Eq.~\eqref{eq:XXZ} and compute 
$\hat{\rho}^\text{DE}_c$ efficiently, we exploit the parity invariance of $\hat{H}$ 
to work in either the even or the odd sector. Since $[\hat{H},\hat{S}^z]=0$, where 
$\hat{S}^z=(\sum_i \sigma_i^z)/2$, we also diagonalize each $S^z$ sector independently. 
The latter results in another major 
advantage of using an NLCE tailored for the initial ground state. Whereas for 
finite-temperature NLCEs all $S^z$ sectors need to be diagonalized, 
for ground-state NLCEs only the $S^z$ sector (or sectors) that contains the initial 
state need to be diagonalized.

For the $XXZ$ model, which only has nearest-neighbor interactions, there is 
one cluster (with $l$ contiguous sites) in the $l^\text{th}$ order of the NLCE. 
For that cluster, $\hat{\rho}^\text{DE}_c$ [Eq.~\eqref{eq:rhode}] 
only needs to be computed in the following two sectors: (i) if $l$ is even, 
the ground states of the AF Ising chain are in the $S^z=0$ sector, so only that sector 
needs to be considered. Within the $S^z=0$ sector, one has to consider both the even 
($e$) parity sector, for which $W^{c,e}_\alpha=|\langle\alpha^e_c|a^e_c\rangle|^2$ 
with $|a^e_c\rangle=(|\ldots\uparrow\downarrow\uparrow\downarrow\ldots\rangle
+|\ldots\downarrow\uparrow\downarrow\uparrow\ldots\rangle)/\sqrt{2}$,
and the odd ($o$) parity sector, for which $W^{c,o}_\alpha=|\langle\alpha^o_c|a^o_c\rangle|^2$ 
with $|a^o_c\rangle=(|\ldots\uparrow\downarrow\uparrow\downarrow\ldots\rangle
-|\ldots\downarrow\uparrow\downarrow\uparrow\ldots\rangle)/\sqrt{2}$.
Note that $|a^e_c\rangle$ ($|a^o_c\rangle$)
is the even (odd) parity ground state of the AF Ising chain, 
while $|\alpha^e_c\rangle$ ($|\alpha^o_c\rangle$) are the even (odd) parity 
eigenstates of the $XXZ$ Hamiltonian. (ii) If $l$ is odd, the ground states of the AF Ising chain
are in the $S^z=1/2$ and $S^z=-1/2$ sectors. In each of those sectors, 
one only needs to consider the one with even parity. For $S^z=1/2$, one needs to calculate 
$W^{c,e}_\alpha=|\langle\alpha^e_c|a^e_c\rangle|^2$, with 
$|a^e_c\rangle=|\ldots\uparrow\downarrow\uparrow\downarrow\uparrow\ldots\rangle$
being one of the ground states of the AF Ising chain when $l$ is odd,
and, for $S^z=-1/2$, one needs to calculate 
$W^{c,e}_\alpha=|\langle\alpha^e_c|a^e_c\rangle|^2$ with 
$|a^e_c\rangle=|\ldots\downarrow\uparrow\downarrow\uparrow\downarrow\ldots\rangle$
being the other ground state of the AF Ising chain when $l$ is odd. 
$|\alpha^e_c\rangle$ are the even parity eigenstates of the $XXZ$ Hamiltonian
in the corresponding $S^z$ sector. Our calculations are further simplified by 
the fact that the $|a^e_c\rangle$'s and $|\alpha^e_c\rangle$'s for $S^z=1/2$ and $S^z=-1/2$  
are trivially related by a $\uparrow\,\rightleftarrows\,\downarrow$ transformation. 
The procedure we have discussed can be straightforwardly extended to consider 
ground states of other Hamiltonians or other specific pure states. 

\begin{figure*}[!t]
    \includegraphics[width=0.9\textwidth]{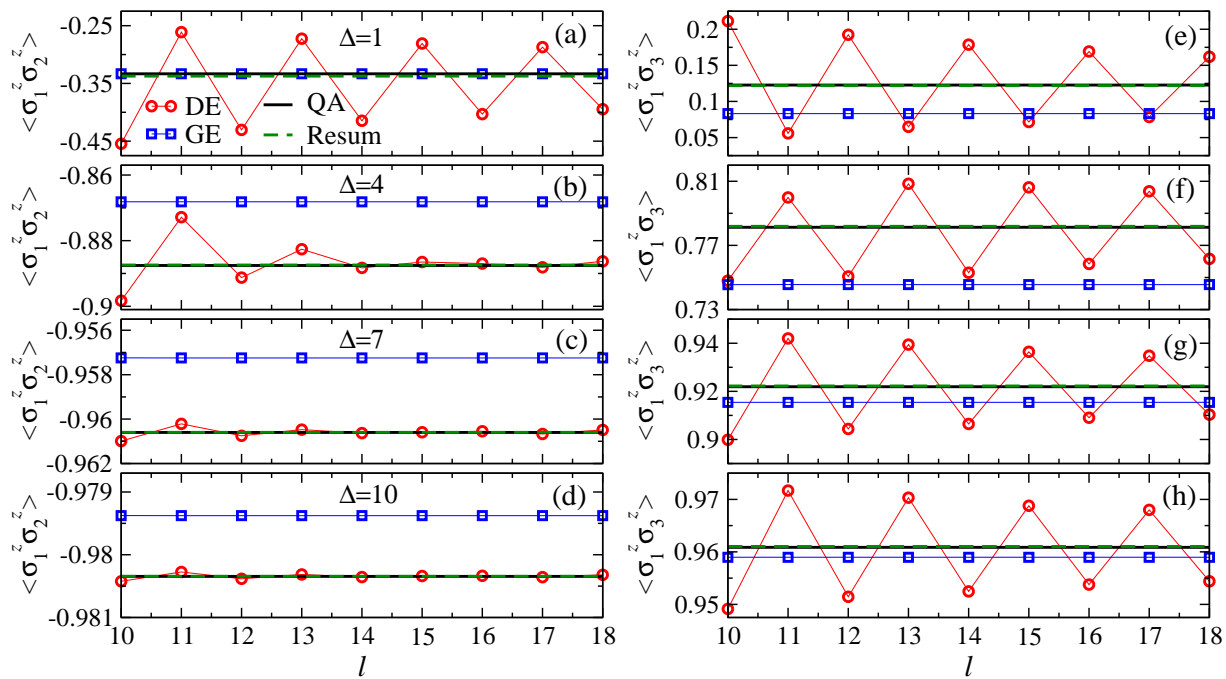}
\vspace{-0.1cm}
\caption{(Color online) Nearest [(a)--(d)] and next-nearest [(e)--(h)] neighbor $\sigma^z\sigma^z$
correlations for quenches with $\Delta=1$ [(a),(e)], $\Delta=4$ [(b),(f)], \
$\Delta=7$ [(c),(g)], and $\Delta=10$ [(d),(h)]. We show results for the last nine
orders of the NLCE for the diagonal ensemble (DE) and for the grand-canonical ensemble (GE).
Lines joining the symbols are provided to guide the eye. 
For all $\Delta$'s, the GE results are essentially converged for all orders shown 
here. The horizontal continuous lines are the results from the quantum action 
method (QA) \cite{wouters_brockmann_14,pozsgay_mestyan_14}, 
and the horizontal dashed lines are the NLCE results after resummations (see text) 
using Wynn's algorithm (Resum). The results of the resummations, and of the NLCE 
bare sums when converged, are virtually indistinguishable 
from those of the quantum action method. The GE results, on the other hand, are clearly 
different except for $\langle\sigma^z_1\sigma^z_2\rangle$ when $\Delta=1$ (see text).
}\label{fig:spincorr}
\end{figure*}

As in Ref.~\cite{rigol_14_91}, here we perform a NLCE for observables in the DE
considering clusters with up to 18 sites. For 18 sites, the sector with $S_z=0$
(the largest one) has 48620 states. Using parity, it is split into the even and odd sectors
that have each 24310 states. Those are the largest ones in which the $XXZ$ Hamiltonian 
needs to be diagonalized. Since, (i) we do not need to diagonalize the initial Hamiltonian to obtain the 
ground state (which we know), (ii) we only need to diagonalize the sectors of the final Hamiltonian 
discussed previously, and (iii) the calculation of $W^{c}_\alpha$ is computationally trivial,
our computation times are greatly reduced from those in Ref.~\cite{rigol_14_91}.
In what follows, we denote as $\mathcal{O}^\text{ens}_l$ (the superscript ``ens'' 
stands for the ensemble used) the result obtained for an observable $\mathcal{O}$  
when adding the contribution of all clusters with up to $l$ sites.

In Fig.~\ref{fig:spincorr}, we show results for nearest [(a)--(d)]
and next-nearest [(e)--(h)] neighbor $\sigma^z\sigma^z$ correlations 
as obtained using NLCEs for the DE. Results are reported for quenches with different 
values of $\Delta$ and for $l$ between 10 and 18. For $\Delta=1$, 
$\langle\sigma^z_1\sigma^z_2\rangle^\text{DE}_l$ [Fig.~\ref{fig:spincorr}(a)]
oscillates for even and odd values of $l$, but the 
amplitude of the oscillation decreases with increasing $l$. This suggests that, 
for larger clusters than the ones considered here, the series
converges. With increasing $\Delta$, Figs.~\ref{fig:spincorr}(b)--\ref{fig:spincorr}(d),
one can see that the amplitudes of the oscillations of 
$\langle\sigma^z_1\sigma^z_2\rangle^\text{DE}_l$ decrease, and 
(within the scale of the plots) the results appear converged. The results for 
$\langle\sigma^z_1\sigma^z_3\rangle^\text{DE}_l$ 
[Figs.~\ref{fig:spincorr}(e)--\ref{fig:spincorr}(h)] are qualitatively 
similar to those for $\langle\sigma^z_1\sigma^z_2\rangle^\text{DE}_l$, except that  
convergence does not appear to be achieved (oscillations are visible) for the values 
of $\Delta$ reported. As expected, with increasing nonlocality larger clusters are required to 
achieve convergence. However, as for $\langle\sigma^z_1\sigma^z_2\rangle^\text{DE}_l$, 
the ratio between the amplitude of the oscillations of 
$\langle\sigma^z_1\sigma^z_3\rangle^\text{DE}_l$ and its mean value generally 
decreases as $\Delta$ increases. Hence, the convergence of the NLCE calculations 
improves as $\Delta$ increases. This results from the fact that
the ground state of the final (gapped) Hamiltonian approaches the initial (trivial) state.

The results for $\langle\sigma^z_1\sigma^z_2\rangle^\text{DE}_l$ 
and $\langle\sigma^z_1\sigma^z_3\rangle^\text{DE}_l$
in Fig.~\ref{fig:spincorr} exemplify the possible outcomes of a NLCE. In some instances,
results for an observable converge to a desired accuracy within the cluster sizes accessible 
in the calculations [e.g., Figs.~\ref{fig:spincorr}(b)--\ref{fig:spincorr}(d)]
and in others they do not [e.g., Figs.~\ref{fig:spincorr}(a),
\ref{fig:spincorr}(e)--\ref{fig:spincorr}(h)]. In the former case, the results of the 
bare NLCE sums are all one needs. This was the case in Ref.~\cite{rigol_14_91} for the initial 
temperatures selected in the quenches studied. On the other hand, if the bare NLCE sums 
do not converge to a desired accuracy, one can use resummation
techniques to accelerate convergence and improve accuracy. Useful resummation techniques 
that have been implemented in the context of NLCEs can be found
in Ref.~\cite{rigol_bryant_07_30}. Two of them, Wynn's and Brezinski's algorithms, 
provide particularly accurate results for our series. In a ``cycle'' of these algorithms, 
a series for an observable ($\mathcal{O}^\text{DE}_l$, with $l=1,\ldots,18$ in 
our case) is transformed into a different series with fewer elements. Each cycle is 
expected to improve convergence, with the last element converging to the thermodynamic 
limit result, but can also lead to numerical instabilities. 
We find that, after one cycle, the last elements provided by both algorithms are very similar 
to each other and representative of the outcome of the resummations (except for 
$\langle\sigma^z_1\sigma^z_2\rangle$ when $\Delta=1$ for which 5 cycles are required). 
In Fig.~\ref{fig:spincorr}, we report Wynn's algorithm results 
for the correlation functions (horizontal dashed lines).

In order to gauge the accuracy of the NLCE bare sums and resummations, we 
compare our results to recent analytic ones for 
$\langle\sigma^z_1\sigma^z_2\rangle$ \cite{wouters_brockmann_14} and 
$\langle\sigma^z_1\sigma^z_3\rangle$ \cite{pozsgay_mestyan_14} obtained 
within the quench action method \cite{caux_essler_13,nardis_wouters_14}. 
The latter are depicted in Fig.~\ref{fig:spincorr} as continuous horizontal lines. 
Within the scales in the plots, the NLCE results after 
resummations are virtually indistinguishable from the quench action results.  
The same is true when the NLCE bare sums appear converged, for which the results 
are indistinguishable from the resummed and the quench action ones.

After a quench in integrable systems, such as the $XXZ$ chain 
\cite{cazalilla_citro_review_11_64} studied here, observables are expected to relax 
to the predictions of a generalized Gibbs ensemble (GGE) \cite{rigol_dunjko_07_27}, 
which maximizes the entropy \cite{jaynes_57a,jaynes_57b} given the constraints imposed 
by the conserved quantities that make the system integrable. This has been shown 
to occur in numerical and analytical studies of integrable models that are 
mappable to noninteracting ones \cite{rigol_dunjko_07_27,rigol_muramatsu_06_26,
cazalilla_06,kollar_eckstein_08,iucci_cazalilla_09,fioretto_mussardo_10,
iucci_cazalilla_10,cassidy_clark_11_56,calabrese_essler_11,gramsch_rigol_12_78,
cazalilla_iucci_12,calabrese_essler_12b,essler_evangelisti_12,collura_sotiriadis_13,
caux_essler_13,fagotti_13,fagotti_essler_13a}, where the conserved quantities have been taken to be 
either the occupation of the single-particle eigenstates of the noninteracting model 
or local quantities. In Refs.~\cite{wouters_brockmann_14,pozsgay_mestyan_14},
it was shown that the results from the quantum action method (expected to predict 
the outcome of the relaxation dynamics) and from the GGE based on known local 
conserved quantities are different for quenches in the $XXZ$ chain. 
This has opened a debate as to which other conserved quantities, if any, 
should be included in the GGE so that it can describe observables after relaxation
\cite{fagotti_essler_13,fagotti_collura_14,mierzejewski_prelovsek_14,
goldstein_andrei_14,pereira_pasquier_14,pozsgay_14}.

For the quenches studied here, the differences between the quantum action 
method and the GGE are so small that, except for 
$\langle\sigma^z_1\sigma^z_3\rangle$ close to the Heisenberg point 
\cite{wouters_brockmann_14}, they cannot be resolved within our NLCEs. 
For example, (i) for $\Delta=7$ we obtain from NLCE after resummations that 
$\langle\sigma^z_1\sigma^z_2\rangle^\text{NLCE}=-0.96060$ while the quantum 
action predicts $\langle\sigma^z_1\sigma^z_2\rangle^\text{QA}=-0.960601$
and the GGE predicts $\langle\sigma^z_1\sigma^z_2\rangle^\text{GGE}=-0.960597$; and
(ii) for $\Delta=10$ we obtain from NLCE after resummations that 
$\langle\sigma^z_1\sigma^z_2\rangle^\text{NLCE}=-0.980344$ while the quantum 
action predicts $\langle\sigma^z_1\sigma^z_2\rangle^\text{QA}=-0.9803452$
and the GGE predicts $\langle\sigma^z_1\sigma^z_2\rangle^\text{GGE}=-0.9803447$.

\begin{figure}[!t]
    \includegraphics[width=0.44\textwidth]{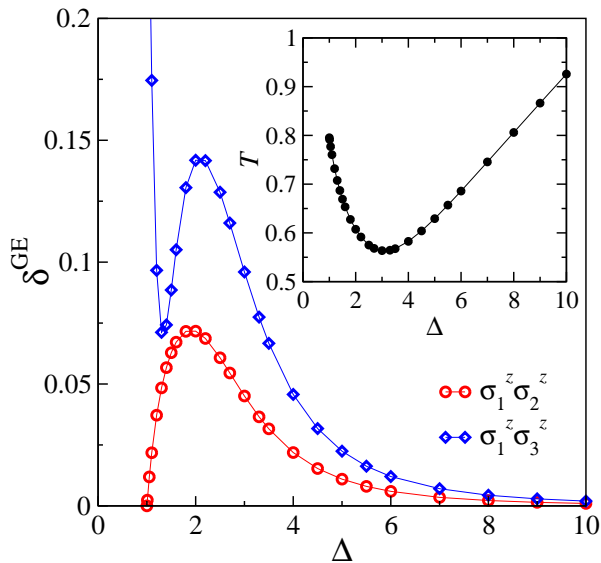}
\vspace{-0.1cm}
\caption{(Color online) Relative differences $\delta_{\sigma^z_1\sigma^z_2}$ and 
$\delta_{\sigma^z_1\sigma^z_3}$ between the GE and the quantum action 
method vs $\Delta$. The differences are seen to go to zero as $\Delta\rightarrow1$
and $\Delta\rightarrow\infty$ and peak for $\Delta\approx2.0$. (Inset)
Effective temperature of the GE vs $\Delta$.
}\label{fig:DiffGEvsQA}
\end{figure}

An important question for experiments is whether the differences 
between the results after equilibration following a quantum quench in an integrable
system and the GE results are large enough that they can be resolved. 
This would allow experimentalists to prove that standard 
statistical mechanics ensembles are unable to describe observables in interacting 
integrable systems after relaxation. In order to address this question, we also 
compute the GE predictions for nearest- and next-nearest-neighbor 
$\sigma^z\sigma^z$ correlations, which we denote as 
$\langle\sigma^z_1\sigma^z_2\rangle^\text{GE}$ and
$\langle\sigma^z_1\sigma^z_3\rangle^\text{GE}$, respectively. We impose 
that the GE must have the same mean energy ($E^\text{GE}$) and expectation value 
of $\hat{S}_z$ ($\langle \hat{S}^z\rangle^\text{GE}$), per site, as the state 
after the quench. Given that the $XXZ$ Hamiltonian is unchanged under a transformation
$\uparrow\,\rightleftarrows\,\downarrow$, a final chemical potential
$\mu=0$ ensures that $\langle \hat{S}^z\rangle^\text{GE}=0$ as in our initial state.
Hence, given the energy per site after the quench ($E^\text{DE}$), all we need 
is to find the temperature $T$ at which $E^\text{GE}=E^\text{DE}$.
We compute $T$ by requiring that the normalized [as in Eq.~\eqref{eq:norm}] 
energy difference between $E^\text{DE}_{18}$ and $E^\text{GE}_{18}$ 
is smaller than $10^{-11}$. We should stress 
that, in all our calculations, $E^\text{DE}_{18}$ and $E^\text{GE}_{18}$ are 
fully converged within machine precision.

In the inset in Fig.~\ref{fig:DiffGEvsQA}, we plot $T$ versus $\Delta$. That plot 
shows that $T$ decreases as $\Delta$ departs from $1$, reaches a minimum near 
$\Delta\approx3$, and then increases almost linearly with $\Delta$ for large values
of $\Delta$. The latter behavior is the result of the increase of the ground-state 
gap with increasing $\Delta$, and the fact that the initial state is not an eigenstate of 
Hamiltonian Eq.~\eqref{eq:XXZ} for any finite value of $\Delta$. This behavior is 
qualitatively similar to the one seen in quenches in the Bose-Hubbard model 
when the initial state is a Fock state with one particle per site \cite{sorg_vidmar_14}. 

Results for $\langle\sigma^z_1\sigma^z_2\rangle^\text{GE}$ 
and $\langle\sigma^z_1\sigma^z_3\rangle^\text{GE}$
versus $l$ are plotted in Fig.~\ref{fig:spincorr}. In all cases one can see
that, for the values of $l$ reported, the NLCE results for the GE are converged within 
the scale of the plots. Furthermore, they are clearly different from the 
results after the quench in all cases but for $\langle\sigma^z_1\sigma^z_2\rangle$ and 
$\Delta=1$. At the Heisenberg point, the $SU(2)$ symmetry of the model results in 
$\langle\sigma^z_1\sigma^z_2\rangle^\text{QA}=\langle\sigma^z_1\sigma^z_2\rangle^\text{GGE}
=\langle\sigma^z_1\sigma^z_2\rangle^\text{GE}$.

In order to quantify the differences between the results after relaxation following 
the quench and the GE predictions, we compute the normalized differences
\begin{equation}\label{eq:norm}
 \delta^\text{GE}_{\sigma^z_i\sigma^z_j}=
 \frac{|\langle\sigma^z_i\sigma^z_j\rangle^\text{GE}-
        \langle\sigma^z_i\sigma^z_j\rangle^\text{QA}|}
       {|\langle\sigma^z_i\sigma^z_j\rangle^\text{QA}|}.
\end{equation}
$\delta^\text{GE}_{\sigma^z_1\sigma^z_2}$ is plotted in the main panel of Fig.~\ref{fig:DiffGEvsQA} 
versus $\Delta$. This quantity first increases as $\Delta$ departs from 1, reaches a 
maximum around $\Delta=2.0$ and then decreases. This is qualitatively similar
to the behavior reported in Ref.~\cite{wouters_brockmann_14} for the differences between 
$\langle\sigma^z_1\sigma^z_2\rangle^\text{GGE}$ and $\langle\sigma^z_1\sigma^z_2\rangle^\text{QA}$. 
There is an important quantitative difference though, $\delta^\text{GE}_{\sigma^z_1\sigma^z_{2}}$ is much 
larger. $\delta^\text{GE}_{\sigma^z_1\sigma^z_3}$ exhibits a qualitatively similar behavior to 
$\delta^\text{GE}_{\sigma^z_1\sigma^z_2}$ except that, for $\Delta\gtrsim1$, it first sharply decreases 
(from $\delta^\text{GE}_{\sigma^z_1\sigma^z_3}=0.32$ for $\Delta=1$) before increasing as 
$\delta^\text{GE}_{\sigma^z_1\sigma^z_2}$ does for $\Delta\gtrsim1$. We note that, for all values of 
$\Delta>1$, the ``memory'' of the initial state (due to integrability) leads to 
$|\langle\sigma^z_1\sigma^z_{2,3}\rangle^\text{QA}|>|\langle\sigma^z_1\sigma^z_{2,3}\rangle^\text{GE}|$. 
The large values attained by $\delta^\text{GE}_{\sigma^z_1\sigma^z_{2}}$ and 
$\delta^\text{GE}_{\sigma^z_1\sigma^z_{3}}$ make them potentially accessible to experimental verification.

In summary, we have shown that NLCEs for the DE, recently introduced in Ref.~\cite{rigol_14_91}, 
can be used to study quenches starting from ground states or other engineered initial states 
of interest. Here we have studied the particular case of quenches in the (integrable) $XXZ$ chain 
starting from the ground state of the AF Ising chain. Our bare NLCE sums (when converged), 
and the results after resummations, were shown to be in excellent agreement with analytic 
results in the thermodynamic limit. Furthermore, we have shown that 
the differences between the outcome of the relaxation dynamics for $\langle\sigma^z_1\sigma^z_2\rangle$
and $\langle\sigma^z_1\sigma^z_3\rangle$ in such quenches and the thermal predictions 
are large enough that they could potentially be resolved in experiments.

\begin{acknowledgments}
This work was supported by the U.S. Office of Naval Research. We are grateful to 
J. S. Caux and M. Brockmann for stimulating discussions, to J. De Nardis for 
providing all quench action and GGE results reported in this manuscript, and 
to D. Iyer, E. Khatami, and R. Mondaini for critical reading of the manuscript.
\end{acknowledgments}

\end{document}